\documentstyle[12pt]{article}

 2
\def\BI{{\rm 1\!l}}
\def\pois#1#2{\{#1,#2\}}
\def\Tr{\hbox{{\rm Tr}}}
\def\Eq#1{{\begin{equation} #1 \end{equation}}}
\def\blackbxx{\hbox{\leaders\hrule width 0.6em height 0.6em\hskip0.6em}}
\def\boxit{{\vbox{\hrule\hbox{\vrule
\vbox{{\phantom{\blackbxx}}
 }\vrule}\hrule}}\kern0.25em}

\newcommand{\ba}{\begin{eqnarray}}
\newcommand{\ea}{\end{eqnarray}}
\def\dmnone{\matrix{{}\cr d^{(-)}_n \cr {}^1}}
\def\dmmtwo{\matrix{{}\cr d^{(-)}_m \cr {}^2}}

\def\dpnone{\matrix{{}\cr d^{(+)}_n \cr {}^1}}

\def\dpmtwo{\matrix{{}\cr d^{(+)}_m \cr {}^2}}

\def\dpmmtwo{\matrix{{}\cr d^{(\pm)}_m \cr {}^2}}

\def\Bnone{\matrix{{}\cr B_n \cr {}^1}}
\def\Bmtwo{\matrix{{}\cr B_m \cr {}^2}}
\def\Gnone{\matrix{{}\cr G_n \cr {}^1}}
\def\Gmtwo{\matrix{{}\cr G_m \cr {}^2}}
\def\Enone{\matrix{{}\cr E_n \cr {}^1}}
\def\Emtwo{\matrix{{}\cr E_m \cr {}^2}}

\def\lpmnone{\matrix{{}\cr \ell^{(\pm)}_n \cr {}^1}}
\def\lpmmtwo{\matrix{{}\cr \ell^{(\pm)}_m \cr {}^2}}
\def\lpnone{\matrix{{}\cr \ell^{(+)}_n \cr {}^1}}
\def\lmnone{\matrix{{}\cr \ell^{(-)}_n \cr {}^1}}
\def\lmmtwo{\matrix{{}\cr \ell^{(-)}_m \cr {}^2}}
\def\jnone{\matrix{{}\cr  j(x_n) \cr {}^1}}
\def\jmtwo{\matrix{{}\cr j(x_m) \cr {}^2}}
\def\jxone{\matrix{{}\cr  j(x) \cr {}^1}}
\def\jytwo{\matrix{{}\cr j(y) \cr {}^2}}

\def\vmnone{\matrix{{}\cr v^{(-)}_n \cr {}^1}}
\def\vmmtwo{\matrix{{}\cr v^{(-)}_m \cr {}^2}}
\def\wmnone{\matrix{{}\cr w^{(-)}_n \cr {}^1}}
\def\wmmtwo{\matrix{{}\cr w^{(-)}_m \cr {}^2}}
\def\wpmtwo{\matrix{{}\cr w^{(+)}_m \cr {}^2}}
\def\knone{\matrix{{}\cr  k^{(-)}_n \cr {}^1}}
\def\kmtwo{\matrix{{}\cr  k^{(-)}_m \cr {}^2}}
\def\kpnone{\matrix{{}\cr  k^{(+)}_n \cr {}^1}}
\def\vpmtwo{\matrix{{}\cr v^{(+)}_m \cr {}^2}}
\def\ktnone{\matrix{{}\cr \tilde k^{(-)}_n \cr {}^1}}
\def\ktmtwo{\matrix{{}\cr \tilde k^{(-)}_m \cr {}^2}}
\begin{document}
\begin{flushright}
UAHEP 9612\\
DSFNA-T-57-96\\
December 1996\\
\end{flushright}

\centerline{ \LARGE Hidden Quantum Group Symmetry}
\centerline{ \LARGE in the Chiral Model}
\vskip 2cm

\centerline{ {\sc  A. Stern* and P. Vitale**}}

\medskip

\centerline{* Dept. of Physics and Astronomy, Univ. of Alabama,
Tuscaloosa, Al 35487, U.S.A.}
\medskip

\centerline{**  Dip.to di Scienze Fisiche, Universit\`a di Napoli,
80125 Napoli, Italy}

\vskip 2cm

\vspace*{5mm}

\normalsize
\centerline{\bf ABSTRACT}
We apply the $SL(2,C)$ lattice Kac-Moody algebra of
 Alekseev,  Faddeev and  Semenov-Tian-Shansky to obtain a new
lattice description of the $SU(2)$ chiral model in two dimensions.
  The system has
a global quantum group symmetry and it can be regarded as a deformation
of two different theories.
 One is the nonabelian Toda lattice which
is obtained in the limit of infinite central charge, while the other
is a nonstandard Hamiltonian description of the
chiral model obtained in the continuum limit.

\newpage
\scrollmode
\section{Introduction}

The two dimensional chiral model is an example of a field
theory which has an infinite number of conserved charges, yet their
quantization is problematic.  The reason is due to the theory being
`nonultralocal', which means that Schwinger-like terms appear in the
algebra of the Lax matrices.
  This in turn leads to difficulties in
defining the Poisson brackets of the conserved charges
constructed out of
the monodromy     matrices.

   A number of proposal for removing the above ambiguities
in dealing with nonultralocal models  have been made.
 \cite{Ma},\cite{fm},\cite{Hl},\cite{STSS},\cite{BB},\cite{RSV2}
  One which we wish to pursue here is to put the theory
 on the lattice.  (Our lattice will be one dimensional, with time
  remaining  continuous.)  Here we shall examine two different
schemes for descretizing the $SU(2)$ chiral model.
  The first which we shall study is already
known and it is based on the nonabelian Toda lattice.\cite{fm}
It leads to the standard Hamiltonian formalism in the continuum limit,
and the classical theory  has a global canonical symmetry.
  The second is new and it is based on
 the  lattice Kac-Moody
 algebra of Alekseev, Faddeev and Semenov-Tian-Shansky.\cite{afs}
 (also see \cite{STS},\cite{RTS},\cite{FV})
The continuum limit of this theory gives a nonstandard Hamiltonian
 formulation of the $SU(2)$ chiral model found by Rajeev which
 is based on the $SL(2,C)$ current algebra.
 \cite{Raj},\cite {RSV},\cite{RSV2},\cite{aS}  The
description (both on the lattice and in the continuum limit)
is canonically inequivalent to the previous one.
Rather, it is a deformation of the previous description.
The classical lattice Kac-Moody algebra has a
 global Lie-Poisson symmetry (the classical counterpart of a quantum
group symmetry \cite{PC})
and it is characterized by a central charge $\kappa$.
Upon taking the limit of infinite central charge one recovers
the previous description based on the nonabelian Toda lattice.

A quantum group symmetry was shown previously to exist in the
WZNW model.\cite{als}  Here we conclude that a quantum group
symmetry can be present in a theory that contains no
 Wess-Zumino term.
We shall not investigate the question of
conserved charges and integrability for the lattice
theories in this article, but intend it to be the subject for   a
future article.

We begin in section 2
by reviewing the continuum theory of the chiral model.
The discretization of the standard Hamiltonian formalism is given
in section 3.  In section 4 we
write down the classical version of the lattice algebra of Alekseev,
Faddeev and Semenov-Tian-Shansky.  We then examine the * operation
for this system and find a consistent algebra only for the
case of a real central charge.  After confirming that its
continuum limit is the   $SL(2,C)$ Kac-Moody algebra,
 we show that in the limit of
infinite central charge the lattice algebra coincides with that of the
nonabelian Toda lattice.  A  Lie-Poisson gauge
transformation is  revealed
for the lattice algebra in section 5.
   In section 6 we specify a Hamiltonian
for this system which agrees with ref. \cite{Raj}
 in the continuum limit
and the Toda lattice Hamiltonian in the limit of infinite central charge.
The Hamiltonian breaks the local Lie-Poisson symmetry of the Poisson
brackets to a global one.
In section 7 we consider restoring the local symmetry in order to
write a Lie-Poisson lattice gauge theory, and in so doing we
recover previously
found systems \cite{fro},\cite{BSV}.  Here we shall also show
how the non local algebra of ref. \cite{afs} can be expressed
entirely in terms of a local algebra and that the latter is
related to the algebra of the classical double.
Some concluding remarks are  made in section 8.

\section{ The Continuum Theory}

In two dimensions, the $SU(2)$
chiral model dynamics can be specified in terms
of the currents $I_\alpha$ and $J_\alpha$, $\alpha=1,2,3$, the
equations of motion being
\ba \dot I_\alpha &=& J'_\alpha     \cr
    \dot J_\alpha &=& I'_\alpha  -\epsilon_{\alpha\beta\gamma}I_\beta
     J_\gamma \;,\label{nem}\ea
     where  the dot denotes the time derivative,
the prime denotes the space derivative, and
$\epsilon_{\alpha\beta\gamma}$ are the structure constants for $SU(2)$.
 Here and throughout this paper we specialize to the case of $SU(2)$.
The equations can be generalized to include effects of a
Wess-Zumino term, but we shall not consider such a modification here.
(For a discussion see section 7.)

We shall be concerned with
two canonically inequivalent Hamiltonian descriptions of the
chiral model, which we refer to as the `standard' and
`alternative' formulations.    In the standard Hamiltonian
 formulation the equations of motion (\ref{nem})
 are recovered from the following
Hamiltonian and Poisson brackets:
\Eq{      H_{0}=-\frac{1}{4\chi^{2}}\int dx\;(I_\alpha I_\alpha
+J_\alpha J_\alpha ) \;,\label{ham1}        }
\ba
 \frac{1}{2 \chi^{2}} \{ I_{\alpha}(x) ,I_{\beta}(y)   \}_{0}
&=&      \varepsilon_{\alpha \beta \gamma} I_{\gamma}(x)\delta (x-y)  \;,
\cr \frac{1}{2 \chi^{2}}\{ I_{\alpha}(x) ,J_{\beta}(y) \}_{0}
&=&     \varepsilon_{\alpha \beta \gamma} J_{\gamma}(x)
     \delta (x-y) -\delta_{\alpha \beta} \delta^{\prime}(x-y)~, \cr
 \frac{1}{2 \chi^{2}}\{ J_{\alpha}(x) ,J_{\beta}(y) \}_{0}
&=& 0~,\label{PBJJ}   \ea
$\chi$ being an arbitrary constant.

The alternative canonical formalism  for the
chiral model was introduced in refs. \cite{Raj},\cite{RSV},\cite{RSV2}
which replaces  (\ref{ham1})  and
(\ref{PBJJ}) by\footnote{In comparing with
 ref. \cite {RSV}, the parameters $\rho$ and $\tau$ of
that reference are given by $\rho=0$ and $\tau^2=-\xi^{-2}$.}
\Eq{      H_{1/\xi}=-\frac{1}{4\chi^{2}\;(1+\xi^{-2})^2}
\int dx\;(I_\alpha I_\alpha  +J_\alpha J_\alpha ) \label{ham2} }
\ba
 \frac{1}{2 \chi^{2}} \{ I_{\alpha}(x) ,I_{\beta}(y)   \}_{1/ \xi}
&=& (1+\xi^{-2})\;
 \varepsilon_{\alpha \beta \gamma} I_{\gamma}(x)\delta    (x-y)   \;,\cr
 \frac{1}{2 \chi^{2}}\{ I_{\alpha}(x) ,J_{\beta}(y) \}_{1/ \xi}
&=& (1+\xi^{-2})  \;
 \varepsilon_{\alpha \beta \gamma} J_{\gamma}(x)
     \delta (x-y) -      (1+\xi^{-2})^2   \;
     \delta_{\alpha \beta} \delta^{\prime}(x-y) ,  \cr
 \frac{1}{2 \chi^{2}}\{ J_{\alpha}(x) ,J_{\beta}(y) \}_{1/ \xi}
&=& -{ \xi^{-2}} (1+\xi^{-2}) \;
 \varepsilon_{\alpha \beta \gamma} I_{\gamma}(x)\delta    (x-y)   \;,
\label{PBJJ2}    \ea
for a real constant $\xi$.   For finite $\xi$ the Poisson structure
(\ref{PBJJ2}) is canonically   inequivalent to
(\ref{PBJJ}).  On the other hand, the Hamiltonian (\ref{ham2})
 and Poisson structure (\ref{PBJJ2})  reduces to
  (\ref{ham1})  and  (\ref{PBJJ})  when $\xi\rightarrow \infty$.
Therefore (\ref{ham2}) and (\ref{PBJJ2})
 give a one-parameter deformation of the standard
canonical formalism.

The Poisson bracket algebra (\ref{PBJJ2}) is equivalent to
the $SL(2,C)$ Kac-Moody algebra.   To see this we can write
 \Eq{I_\alpha = 2\chi^2
     (1+ \xi^{-2}) \tilde I_\alpha  \;,\quad
  J_\alpha = -{{2 \chi^2}\over \xi }  (1+ \xi^{-2}) \tilde J_\alpha\;,
 \label{IItJJt}}    and then (\ref{PBJJ2}) in terms of the currents
$ \tilde I_\alpha(x)$ and $ \tilde J_\alpha(x)$   becomes
 \ba \{ \tilde I_\alpha(x), \tilde I_\beta(y)\}
  &=&-\{\tilde J_\alpha(x), \tilde J_\beta(y)\}
  =\epsilon_{\alpha \beta \gamma} \tilde I_\gamma(x) \delta(x-y)\;,\cr
& & \cr   \{ \tilde I_\alpha(x), \tilde J_\beta(y)\} &=&
\epsilon_{\alpha \beta \gamma} \tilde J_\gamma(x) \delta(x-y)+\frac{
\xi}{2\chi^2}
 \delta_{\alpha\beta}\partial_x \delta(x-y) \;.\label{kma}\ea
Note that   this  is  not the most general
 $SL(2,C)$ Kac-Moody   algebra as a second central term is allowed
in the algebra (see discussion in section 7).

In the section which follows we examine the discretization of the
standard Hamiltonian formalism, while the
 discretization of the alternative  Hamiltonian formalism is given in
sections 4 through 6.

\section{The Nonabelian Toda Lattice}

Here we show that
the lattice version of the standard Hamiltonian description
 defined by
  (\ref{ham1})  and  (\ref{PBJJ})  can be formulated in terms
of the nonabelian Toda lattice.\cite{fm}  Once again we specialize to
the case of $SU(2)$.  Then
this system  can be written   in terms of $2\times 2$
 matrices $G_n$ and $B_n$,
$n=1,2,...,N$, $N$ being the total number of lattice sites, where
 $G_n$  is traceless and antihermitean, while $B_n$
is an $SU(2)$ matrix.  For their  Poisson brackets we take
     \ba  \frac 2{\chi^2}
\pois{\Gnone}{\Gmtwo} &=&  [C,\Gmtwo ] \delta_{n,m} \;,\cr & & \cr
  \frac 2{\chi^2}
\pois{\Gnone}{\Bmtwo} &=&  C\Bmtwo  \delta_{n,m-1} -
\Bmtwo C\delta_{n,m} \;,\cr    \{\Bnone,\Bmtwo\} &=& 0 \;,\label{tlpbp}
\ea
where we utilize tensor product notation.  Here
$\Gnone =G_n\otimes \BI$, $\Gmtwo =\BI \otimes G_m$,
$\Bnone =B_n\otimes \BI$, $\Bmtwo =\BI \otimes G_m$
and $C=\sigma_i \otimes \sigma_i$ is adjoint invariant.
This    algebra is nonlocal due to the  interactions  between
neighboring sites in the second Poisson bracket.  Nevertheless,
it can be reexpressed in terms of a local algebra,
 more specifically, in terms of variables which span the
 product space of cotangent bundles of $SU(2)$.  We show how to do
this in section 7.

The Poisson structure  defined by the brackets  (\ref{PBJJ})
  is the  continuum limit of the  nonabelian Toda lattice algebra
 (\ref{tlpbp}).   To see this we write    \Eq{
   G_{n}  = -i  a\sigma_\alpha {\cal I}_\alpha (x_n)/2\;,\quad
 B_n = \exp{\{ -ia \sigma_\alpha {\cal J}_\alpha(x_n)/2\}
 } \;,\label{AIBJ}}  $\sigma_\alpha$ being the Pauli matrices,
   and make the identification  of ${\cal I}_\alpha(x) $ and
${\cal J}_\alpha(x) $  with $I_\alpha (x) $ and $J_\alpha (x) $.

Canonical transformations are present  for the Poisson
 brackets    (\ref{tlpbp}).  The latter are preserved under:
\Eq{ G_n \rightarrow G'_n= v_n^{-1} G_n v_n \;,\quad
 B_n \rightarrow B'_n =
 v_{n-1}^{-1} B_n v_n \;,\quad v_n \in SU(2) \;.\label{ct}}
  These transformations correspond to $SU(2)$ gauge (or local)
  symmetries as we can associate an $SU(2)$
   group element $v_n$ with each  link on the lattice.
Furthermore from the Poisson brackets    (\ref{tlpbp})
  they are generated by the  set of all $G_n$.

For the lattice dynamics  we need to specify the  Hamiltonian
$ H^{lat}_0$.    We take
   \Eq{ H^{lat}_0 = \frac1{2a\chi^2} \sum_n \Tr\; (G_n^2 +
B_n + B_n^\dagger) \;,\label{tlH}}
from which we recover (up to an infinite constant)
the chiral model Hamiltonian $H_0$ in
the $a\rightarrow 0$ limit.    Regarding the symmetries, we note that
 $\Tr \;B_n $ and hence
$ H^{lat}_0 $ are not invariant under the most
general canonical transformations (\ref{ct})
of the Poisson brackets (\ref{tlpbp})
[although $\Tr \;G_n^2$ is invariant].  Rather, they are preserved
only under the action of the global subgroup, i.e. where (\ref{ct}) is
restricted by $v_1=v_2= ...=v_N=v$.    The generator of this subgroup
is $G=\sum_{n=1}^N G_n$.  It is conserved provided we take the underlying
manifold on which  the lattice  is constructed
to be a circle, whereby the boundary conditions are periodic
$B_{n+N}=B_n$.   This is seen from
  the equations of motion which follow from (\ref{tlH}):
 \ba a\dot G_n &=& \frac12 [B_{n+1}-B_n
-B_{n+1}^\dagger +B_n^\dagger]_{t\ell} \;,   \cr
    a \dot B_n &=& B_nG_n -G_{n-1}B_n\;, \ea
 where   $[A]_{t\ell}$ denotes the traceless part of $2\times 2$ matrix
 $A$, i.e.,  $[A]_{t\ell}=A-\frac 12 \Tr (A) \times \BI$,  $\BI$
denoting the $2\times 2$ unit matrix.  Then
\Eq{ a\dot G= \frac12 [B_{N+1}-B_1
-B_{N+1}^\dagger +B_1^\dagger]_{t\ell} \;,   } which vanishes
after applying the periodic boundary conditions.

The Poisson brackets (\ref{tlpbp}) and Hamiltonian (\ref{tlH})
give the lattice formulation of the standard Hamiltonian description
of the chiral model.  In the next three sections we
develop the lattice formulation of the alternative
Hamiltonian description  of the chiral model.
Since from (\ref{kma}) it is based on the $SL(2,C)$ Kac-Moody algebra
we must address the discretization of this algebra, which is done
in the following section.

\section{Discretization of the $SL(2,C)$ Kac-Moody algebra}

Here we first
write down the classical version of the lattice algebra of Alekseev,
Faddeev and Semenov-Tian-Shansky.\cite{afs}   It is characterized by
a central charge $\kappa$ which we will relate to the parameters
$\xi$ and $\chi$ appearing in the alternative Hamiltonian formulation
of the chiral model.    We  examine the * operation
for this system and find a consistent algebra only for the
case of  real $\kappa$.  We then confirm that
continuum limit agrees with the $SL(2,C)$ Kac-Moody algebra
of the alternative Hamiltonian formalism given by
 (\ref{kma}).   We further
 show that in the limit $\kappa\rightarrow
\infty$   the lattice algebra coincides with that of the
nonabelian Toda lattice (\ref{tlpbp}).

\subsection{The Classical Lattice Algebra}

The classical version of the discretized $SL(2,C)$ Kac-Moody
 algebra\cite{afs}  (also see \cite{STS}, \cite{RTS})
 is given in terms of  $SL(2,C)$ group matrices  $d^{(-)}_n$,
 where $n$ again
 labels the lattice points.  $d^{(-)}_n$ satisfy the Poisson
brackets:                         \Eq{
 \{\dmnone,\dmmtwo\}=-\biggl(\dmnone \dmmtwo  r  +   \tilde r  \dmnone
\dmmtwo\biggr)\delta_{n,m}+  \dmnone r\dmmtwo\delta_{n,m-1}
+\dmmtwo \tilde r \dmnone\delta_{n,m+1}\;. \label{lkm}  }
Here $\dmnone,\dmmtwo$, $r$ and $\tilde r$ denote $4 \times 4$ matrices
with
$\dmnone =d^{(-)}_n\otimes \BI$, $\dmmtwo =\BI \otimes d^{(-)}_m$ and
$r$, $\tilde r$ given by
\Eq{ r={{i}\over {2\kappa
}}\pmatrix{1 & & &  \cr &-1 & &  \cr &4 &-1& \cr
& & & 1\cr} \quad \label{rmat}\;,\quad \tilde r=-r^T\;, }
$T$ denoting transpose.    $\kappa$ is a constant which serves
the role of the central charge which we will relate later to $\xi$
 and $\chi$.   It can be checked that the
Poisson brackets are skew symmetric and satisfy the Jacobi identity.
In addition, $\det{d^{(-)}_n}$ is in the center of the algebra and hence
can be set to unity.
The difference of $r$ and $\tilde r$ is equal to the adjoint invariant,
\Eq{C=-i\kappa(r-\tilde r)\;.}

   Ignoring the interactions between neighboring sites, i.e. the
last two terms in  (\ref{lkm}), the Poisson bracket relations
 for  $d^{(-)}_n$ define the  classical double algebra
 at every site $n$   on the one dimensional lattice.
\cite{D}, \cite{STS}, \cite{maj}, \cite{AM},
\cite{mss}, \cite{af}    However,   due to the interactions
 the full space  is not simply a product
of classical doubles, and it is nonlocal.  In section 7 we shall
show how to write this algebra in terms of a local one and we
show that the latter is related to the classical double algebra.

\subsection{* Operation}

Because $d^{(-)}_n$  are complex matrices,
the relations (\ref{lkm})
are insufficient  for determining the entire algebra.
That is, we must enlarge the algebra to include  the brackets of the
$d^{(-)}_n$'s with their hermitean
conjugates $d^{(-)\dagger}_n$, or equivalently, with   \Eq{
d^{(+)}_n={d^{(-)\dagger}_n}^{-1}\; \;.\label{iod}}
  Properties like the
Jacobi identity and skew symmetry
should remain to be satisfied when we make this enlargement.
In addition, we require that the Poisson structure is preserved under
complex conjugation.  This means that if $\alpha$ and $\beta$ are
any two matrix elements of   $d^{(-)}_n$ or $d^{(+)}_m$, then
\Eq{ {\pois\alpha\beta}^*  =\pois{\alpha ^*}{\beta ^*}  \;.\label{inv}}

We have found  solutions to the above requirements
only in the case of $\kappa$ real, which we now assume.
For the brackets of the  $d^{(-)}_n$'s with $d^{(+)}_n$'s we take the
following:
\Eq{\{\dmnone,\dpmtwo\}=-\biggl(\dmnone \dpmtwo r+  r\dmnone
\dpmtwo\biggr)\delta_{n,m}+\dmnone r\dpmtwo\delta_{n,m-1} +
\dpmtwo  r \dmnone\delta_{n,m+1} \;.\label{dmdp}}
An alternative way to write these Poisson brackets is
\Eq{\{\dpnone,\dmmtwo\}=-\biggl(\dpnone \dmmtwo \tilde r+ \tilde r\dpnone
\dmmtwo\biggr)\delta_{n,m}+\dpnone \tilde r\dmmtwo\delta_{n,m-1} +
\dmmtwo \tilde  r \dpnone\delta_{n,m+1} \;.}   These relations are
obtained from (\ref{dmdp}) using the property of skew symmetry,
 switching the indices $m$ and $n$, as well as
the order of the vector spaces
in the tensor product, $1 \rightleftharpoons 2$.
Upon switching the  order of the vector spaces, $r \rightarrow
r^T$ and $\tilde r \rightarrow   \tilde r^T$.
Then using (\ref{iod}),  \ba
\{\dmnone^\dagger,\dpmtwo^\dagger\}&=&
\{\dpnone^{-1},\dmmtwo^{-1}\}\cr  &=&\dpnone^{-1}\dmmtwo^{-1}
\{\dpnone,\dmmtwo\}  \dpnone^{-1}\dmmtwo^{-1}  \cr &=&
-\biggl(\dmnone^\dagger \dpmtwo^\dagger \tilde r+ \tilde r\dmnone^\dagger
\dpmtwo^\dagger\biggr)\delta_{n,m}\cr & &\quad
 +   \dpmtwo^\dagger \tilde r \dmnone^\dagger\delta_{n,m-1}
+\dmnone^\dagger  \tilde r\dpmtwo^\dagger\delta_{n,m+1}
  \;.\label{dmdpdag} \ea
By comparing (\ref{dmdpdag}) with
(\ref{dmdp}), we see that the property (\ref{inv})  is satisfied and
hence the Poisson structure is preserved under complex conjugation.

 In addition to the relations (\ref{lkm}) and (\ref{dmdp}), we can
obtain the Poisson brackets for
$d^{(+)}_n$ with $d^{(+)}_m$ by taking the complex conjugate of
(\ref{lkm}), again assuming the property (\ref{inv}).  We find \Eq{
 \{\dpnone,\dpmtwo\}=-\biggl(\dpnone \dpmtwo r+ \tilde r \dpnone
\dpmtwo\biggr)\delta_{n,m} + \dpnone \tilde r\dpmtwo\delta_{n,m-1}+
\dpmtwo r \dpnone\delta_{n,m+1} \;. \label{dpdp}  }

The brackets (\ref{lkm}),
 (\ref{dmdp}) and (\ref{dpdp}) completely specify the Poisson structure.
We note that $r$ and $\tilde r$ in the first two terms
can be interchanged in eqs. (\ref{lkm})
 and (\ref{dpdp}), due to $C$ being an adjoint
 invariant.  It can be checked that the complete set of
Poisson brackets  (\ref{lkm}),  (\ref{dmdp}) and (\ref{dpdp})
 are skew symmetric and satisfy the
 Jacobi identity, and that
$\det{d^{(\pm)}_n}$ is in the center of the algebra.

\subsection{The Continuum Limit}

 To recover the $SL(2,C)$ Kac-Moody algebra  in the continuum limit
we set \Eq{d^{(-)}_n=\exp{\{\frac{a}\kappa j(x_n)\}}\;,  \quad
d^{(+)}_n=\exp{\{-\frac{a}\kappa j(x_n)^\dagger\}}\;, \label{dioj}  }
 where $a$ is the lattice
spacing and $j(x_n)$ is the current evaluated at the lattice site $x_n$.
Next we do an expansion in  $a/\kappa$.   From  the Poisson brackets
 (\ref{lkm}), we get
\ba \{\jnone,\jmtwo\}&=&-\frac {\kappa
^2}{a^2}\biggl((r+\tilde r)\delta_{n,m}
-r\delta_{n+1,m} - \tilde r  \delta_{n-1,m} \biggr)  \cr
& &-\frac {\kappa}{2a}\biggl([r,\jnone-\jmtwo]\delta_{n+1,m} -[ \tilde r,
\jnone-\jmtwo]  \delta_{n-1,m} \biggr)\cr
& & +O(1)\;. \ea
Now taking the limit   $a\rightarrow 0$,
 \Eq{ \{\jxone,\jytwo\}=-i[C,\jxone]\delta(x-y)
  +i\kappa C\partial_x \delta(x-y)   \;.\label{jj}}
From   the Poisson brackets (\ref{dmdp}),
\ba \{\jnone,\jmtwo^\dagger\}&=&\frac {\kappa
^2}{a^2}\;r\;(2\delta_{n,m}
-\delta_{n+1,m} -  \delta_{n-1,m})  \cr
& &+\frac {\kappa}{2a}\;[r,\jnone+\jmtwo^\dagger](\delta_{n+1,m}
 -  \delta_{n-1,m})\cr
& & +O(1)\;. \ea
Now taking the limit   $a\rightarrow 0$,
 \Eq{ \{\jxone,\jytwo^\dagger\}=0\;.\label{jjdag}}
Finally after substituting \Eq{j(x)=\tilde I_\alpha(x)\sigma_\alpha +
 i\tilde J_\alpha(x)\sigma_\alpha\;,\quad \xi=\chi^2 \kappa\;,\label{jox}}
where  $\tilde I_\alpha(x)$ and $ \tilde J_\alpha(x)$
are real-valued currents,
 in (\ref{jj}) and (\ref{jjdag}),  we get the $SL(2,C)$ Kac-Moody
algebra  (\ref{kma}).
Here we see that
we obtained only a single central term in the current algebra
due to the restriction of $\kappa$  being real.

Canonical symmetries are  present for the $SL(2,C)$ Kac-Moody
algebra.  The  Poisson brackets  (\ref{kma})
 are preserved under the global $SL(2,C)$ transformations
\Eq{ j(x) \rightarrow j(x)'=
v j(x) v^{-1}  \;,\quad v \in SL(2,C)\;. \label{ctfca} }

\subsection{The Toda Lattice Limit}

The nonabelian Toda lattice algebra (\ref{tlpbp})
 results from the Poisson brackets (\ref{lkm}),
 (\ref{dmdp}) and (\ref{dpdp})  in the limit of infinite central
charge $\kappa$.  Before taking the limit however
 we first parametrize  (locally)
the $SL(2,C)$ group matrices $d_n^{(\pm)}$ in terms of
matrices spanning the subgroups
 $SU(2)$ and $SB(2,C)$ (the Borel group).  We denote them by
  $B_n$ and $\ell_n^{(\pm)}$,
 respectively.  For this purpose we write    \Eq{
  d_n^{(\pm)}=B_n\ell_n^{(\pm)}\;.\label{dBl}}
   From (\ref{iod}) it follows that
\Eq{\ell^{(+)}_n={\ell^{(-)\dagger}_n}^{-1}\; \;.\label{iol}}
The Poisson brackets (\ref{lkm}),  (\ref{dmdp}) and (\ref{dpdp})
 for $d_n^{(\pm)}$ are recovered if we make the
 following choice of brackets for  $B_n$ and $\ell_n^{(\pm)}$
 \ba \{\Bnone,\Bmtwo\} &=& -[r,\Bnone \Bmtwo ]\delta_{n,m} \;,\cr
  \{\lpmnone,\lpmmtwo\} &=& [r,\lpmnone \lpmmtwo ]\delta_{n,m}\;, \cr
  \{\lpnone,\lmmtwo\} &=& [\tilde r,\lpnone \lmmtwo ]\delta_{n,m}\;, \cr
  \{\lmnone,\Bmtwo\} &=&-\Bmtwo r\lmnone\delta_{n,m} +\lmnone r\Bmtwo
  \delta_{n,m-1}\;, \cr
  \{\lpnone,\Bmtwo\} &=&-\Bmtwo \tilde r\lpnone\delta_{n,m}
   +\lpnone \tilde r\Bmtwo \delta_{n,m-1}\;. \label{lbpb}
   \ea   Using (\ref{iol}),
the fifth equation  is the hermitean conjugate of the fourth equation.
Interactions between neighboring sites occur only in these brackets,
once again making this a nonlocal algebra.  In section 7
we show how this algebra can be reexpressed in terms of a local one.

  Next we write \Eq{\ell_n^{(-)}=\exp{\biggl\{
-\frac{2i}\kappa G_{n,\alpha}e^\alpha}\biggr\}\;, \label{lA}}
where $e^\alpha
,\;\alpha=1,2,3$ are the generators of $SB(2,C)$.  We choose the
following representation for them and the $SU(2)$ generators which
we denote by $e_\alpha$:
$$  e_\alpha ={1\over 2} \sigma_\alpha\;,\quad
 e^\alpha ={1\over 2} (i\sigma_\alpha +\epsilon_{\alpha\beta
 3} \sigma_\beta)\;,$$
   In this representation $e^\alpha$, and hence
   $\ell^{(-)}_n$, are lower triangular
  matrices\footnote{  It can be
checked that the Poisson  brackets (\ref{lbpb}) are consistent
with setting the matrix element  $[\ell^{(-)}_n]_{12}=0.$},
and  we can write $r=\frac{2}\kappa e^\alpha \otimes e_\alpha$ and
 $\tilde r=-\frac{2}\kappa e_\alpha \otimes e^\alpha$.  We note here that
the ordering of the $SU(2)$ and $SB(2,C)$ matrices in (\ref{dBl})
appears to be important.  Upon choosing
 $\ell^{(-)}_n$ to  be lower triangular matrices,
we were unable to find a consistent
Poisson structure for   $B_n$ and $\ell_n^{(\pm)}$
 in the case where they appear
 in the reverse order in the definition
 (\ref{dBl}) of $d^{(\pm)}_n$.  On the other hand, the reverse order
is necessary if instead we choose
 $\ell^{(-)}_n$ to  be upper triangular matrices (simultaneously
replacing $e^\alpha$ by upper triangular matrices).

 Now by taking the limit
$\kappa\rightarrow\infty$, the Poisson brackets (\ref{lbpb}) imply
 \ba \{G_{n,\alpha},G_{m,\beta
 }\} &=& \epsilon_{\alpha\beta\gamma}\delta_{n,m} G_{n,\gamma}+O(
 \frac{1}\kappa)\;,\cr
  \{G_{n,\alpha},B_m\} &=&-iB_me_\alpha
   \delta_{n,m} +ie_\alpha B_m \delta_{n,m-1}+O(\frac{1}\kappa)  \;,\cr
 \{\Bnone,\Bmtwo\} &=&  O(\frac{1}\kappa)   \;.\label{tlpb}
  \ea
By defining $G_n=-i\chi^2 G_{n,\alpha} \sigma_\alpha\;,\chi $,
  the zeroth order terms in
 (\ref{tlpb}) can be written   according to (\ref{tlpbp}).
We have thus recovered
 the  algebra of the nonabelian Toda lattice.

We note that the continuum limit of the nonabelian Toda lattice cannot
be described in terms of the currents $\tilde I_\alpha$
and $\tilde J_\alpha$ as the $SL(2,C)$ algebra (\ref{kma})
 is illdefined when $\kappa \rightarrow \infty$.
On the other hand, we note that when $\kappa$ is finite we cannot
identify  ${\cal I}_\alpha(x) $ and ${\cal J}_\alpha(x) $
appearing in (\ref{AIBJ})  with the currents
 $I_\alpha (x) $ and $J_\alpha (x) $ as we did for the nonabelian
Toda lattice.
 From (\ref{IItJJt}) and (\ref{dioj}), $d^{(-)}_n$ can be written
 \Eq{d^{(-)}_n=\exp{\biggl\{ a e_\alpha \; \frac{\frac 1{\chi^2
 \kappa} I_\alpha(x_n ) -i J_\alpha(x_n)  }{1+\frac 1{\chi^4\kappa^2}}
 \biggr\} }        \;, \label{deIJ}  }
while (\ref{dBl}), (\ref{lA}) and (\ref{AIBJ})  imply
 \Eq{d^{(-)}_n=\exp{\{-i a e_\alpha    {\cal J}_\alpha (x_n)\}}
\exp{\{-i \frac a{\chi^2\kappa}e^\alpha {\cal I}_\alpha (x_n)\}} \;.
\label{deJeI}}  By comparing (\ref{deJeI}) and (\ref{deIJ})
 we in general get a complicated expression for the continuum variables
$I_\alpha$ and $J_\alpha$ in terms of
${\cal I}_\alpha$ and ${\cal J}_\alpha$.

\section{Lie-Poisson Symmetry}

We now discuss the symmetry properties  for the above
Poisson structure, i.e. the lattice algebra  defined by
the Poisson brackets (\ref{lkm}),   (\ref{dmdp}) and (\ref{dpdp}).
When   we took two different limits
($a\rightarrow 0$ and $\kappa\rightarrow \infty$)
  of this algebra we obtained Poisson bracket
algebras (\ref{kma}) and (\ref{tlpbp}) which, as we saw,
admit canonical transformations.  They were given by
 (\ref{ct}) for the case of the nonabelian Toda theory and
(\ref{ctfca}) for the case of the Kac-Moody algebra.
Although  the  $a\rightarrow 0$ and $\kappa\rightarrow \infty$ limits
 produce systems with canonical
symmetries, no such canonical symmetries exist for this algebra
 before taking these
limits.  The Poisson bracket algebra defined by  (\ref{lkm}),
   (\ref{dmdp}) and (\ref{dpdp})    instead admits
Lie-Poisson transformations.  They are the classical analogues of
quantum group transformations, and are defined as follows: \cite{PC}

Let ${\cal S}$ denote the space of  symmetries.
Unlike  theories containing canonical symmetries, ${\cal S}$
carries a Poisson structure $\{,\}_{\cal S}$.
Let ${\cal O}$ denote the space of classical observables, which
for us is spanned by the matrices $d^{(\pm)}_n$, whose
 Poisson structure is $\{,\}_{\cal O}$.
${\cal S}$ acting on ${\cal O}$   defines a map,
 $\sigma:{\cal S}\times {\cal O}\rightarrow {\cal O}$.
 ${\cal S}$ induces a Lie-Poisson  action on ${\cal O}$ if
 $\sigma$ is a Poisson map, which means that if $f_1$ and $f_2$ are
functions on  ${\cal O}$, then
 \Eq{ \sigma\circ \{f_1,f_2\}_{\cal O} =
 \{\sigma\circ f_1,\sigma\circ f_2\}_{{\cal O}\times {\cal S}} \;,}
where the product Poisson structure is assumed on
${\cal O}\times {\cal S}$, which means that the symmetry variables
 have zero
Poisson brackets with the classical observables.
[For convenience of notation
we shall drop the subscripts ${\cal S}$ and ${\cal O}$
on the Poisson brackets in the discussion below.]

For the symmetry transformations of the classical
observables $d^{(-)}_n$ we take \Eq{ d^{(-)}_n \rightarrow
 d^{(-)'}_n = w^{(-)}_n d^{(-)}_n v^{(-)}_n \;, \quad
 w^{(-)}_n \;, v^{(-)}_n \in SL(2,C)\;.\label{pm}}
Thus $w^{(-)}_n $ and $ v^{(-)}_n $ span ${\cal S}$ and (\ref{pm})
 defines the map $\sigma$.  The matrices
$w^{(-)}_n $ and $ v^{(-)}_n $  have a nontrivial
Poisson structure.  If we choose   \ba
 \{\vmnone,\vmmtwo\}&=&[  r  ,  \vmnone \vmmtwo]\delta_{n,m} \;,\cr
 \{\wmnone,\wmmtwo\}&=&-[\tilde  r ,\wmnone \wmmtwo]\delta_{n,m}\;, \cr
 \{\vmnone,\wmmtwo\}&=&(\vmnone r\wmmtwo
 -\wmmtwo  r \vmnone)\delta_{n,m-1}\;,   \label{vmvm}\ea
then, as we show below, (\ref{pm}) defines a Poisson map,
 and hence  a Lie-Poisson symmetry.   Here we assume that
$w^{(-)}_n $ and $ v^{(-)}_n $ have zero Poisson brackets with the
observables $ d^{(\pm)}_n $, so  that we get a product Poisson
structure on ${\cal O}\times {\cal S}$.  The first
Poisson bracket in (\ref{vmvm})
 is compatible with group multiplication, which implies that
  $\{ v_n \}$ for every $n$ defines a Lie-Poisson group.
The remaining Poisson brackets in (\ref{vmvm})
 can be obtained from the first
upon setting \Eq{ w_n^{(-)} ={ v_{n-1}^{(-)}}^{-1} \;.\label{wmvm}}
 After making such a restriction we see that the
$d^{(-)}_n $ variables transform analogously to the Toda lattice
variables $B_n$ (\ref{ct}) [only here the symmetry parameters
$v_n^{(-)}$ span $SL(2,C)$ rather than $SU(2)$,
and the transformation is Lie-Poisson rather than canonical].
Since we    can associate a group element $v_n^{(-)}$ with each
 link on the lattice, (\ref{pm})  correspond to gauge transformations.

 To check   that (\ref{pm})
 is a Poisson map we note that the left hand side of (\ref{lkm})
transforms to \ba
 \{\dmnone ',\dmmtwo '\}   &=&\pois {\wmnone\dmnone\vmnone}
 {\wmmtwo\dmmtwo\vmmtwo}\cr & =&
\pois \wmnone\wmmtwo\dmnone\dmmtwo\vmnone\vmmtwo  +
 \wmnone\wmmtwo\dmnone\dmmtwo\pois\vmnone\vmmtwo \cr   & &+
 \wmnone\dmnone\pois\vmnone\wmmtwo\dmmtwo\vmmtwo +
 \wmmtwo\dmmtwo\pois\wmnone\vmmtwo\dmnone\vmnone \cr   & &+
 \wmnone\wmmtwo\pois\dmnone\dmmtwo\vmnone\vmmtwo \;.\ea
Using (\ref{lkm}) and (\ref{vmvm})  we then obtain
 \Eq{   -\biggl(\dmnone ' \dmmtwo ' r  +   \tilde r  \dmnone '
\dmmtwo '\biggr)\delta_{n,m}+  \dmnone ' r\dmmtwo '\delta_{n,m-1}
+\dmmtwo '\tilde r \dmnone '\delta_{n,m+1}\;,   }
which is how the right hand side of (\ref{lkm})
 transforms under (\ref{pm}).  Hence  (\ref{pm}) is a Poisson map.

From the * operation we note that the
$d^{(+)}_n $ variables transform according to
\Eq{ d^{(+)}_n \rightarrow
 d^{(+)'}_n = w^{(+)}_n d^{(+)}_n v^{(+)}_n \;,\label{pmp}}  where
the $(+)$ symmetry parameters (i.e.
$ w^{(+)}_n$ and $ v^{(+)}_n $)  are obtained from
the $(-)$ symmetry parameters (i.e.
$ w^{(-)}_n$ and $ v^{(-)}_n $)  via the operations of
 inverse and conjugations, i.e.
$w^{(+)}_n={w^{(-)\dagger}_n}^{-1}$ and
$v^{(+)}_n={v^{(-)\dagger}_n}^{-1}.$
We can deduce the Poisson structure for all of the variables
$w^{(\pm)}_n$ and $v^{(\pm)}_n$
by again demanding that the transformation
is a Poisson map.  For this we examine how the left and right hand
sides of (\ref{dmdp}) transform under (\ref{pm}) and (\ref{pmp}).
The appropriate Poisson brackets
between the $(-)$ and $(+)$ symmetry parameters are:
   \ba
 \{\vmnone,\vpmtwo\}&=&[  r  ,  \vmnone \vpmtwo]\delta_{n,m} \;,\cr
 \{\wmnone,\wpmtwo\}&=&-[  r ,\wmnone \wpmtwo]\delta_{n,m}\;, \cr
 \{\vmnone,\wpmtwo\}&=&(\vmnone r\wpmtwo
 -\wpmtwo  r \vmnone)\delta_{n,m-1}\;,\cr
 \{\wmnone,\vpmtwo\}&=&(\vpmtwo r\wmnone
 -\wmnone  r \vpmtwo)\delta_{n,m+1}\;,    \label{vmwp} \ea
while the Poisson brackets
between the $(+)$ and $(+)$ symmetry parameters are obtained by taking
the conjugate inverse of (\ref{vmvm})
 and assuming the property (\ref{inv}).
The last three equations in (\ref{vmwp}) follow from the first if we once
again apply (\ref{wmvm}) [which then also implies
$ w_n^{(+)} ={ v_{n-1}^{(+)}}^{-1}$.]

Finally, we remark about the generators of the Lie-Poisson
transformations.  It is in general known that the charges associated
with such transformations are group-valued.\cite{LW}
If we limit our discussion to $SU(2)$ transformations, then
$v^{(-)}_n=v^ {(+)}_n= v_n$ and $d^{(\pm)}_n$ transforms according to
\Eq{ d^{(\pm)}_n \rightarrow   d^{(\pm)'}_n =
 v_{n-1}^{-1}  d^{(\pm)}_n v_n \;, \quad
 v_n \in SU(2)\;,\label{pmsu}}
then its generators are the set of all $SB(2,C)$ matrices
$\ell^{(-)}_n $.  To see this we can compute their
 Poisson brackets with the variables  $d^{(\pm)}_n $.  We find:
\ba
{\lmnone}{}^{-1}\pois\lmnone \dpmmtwo &=& r\dpmmtwo\delta_{n,m-1} -
\dpmmtwo r \delta_{n,m}\cr     &=&\frac 2\kappa e^\alpha\otimes [
e_\alpha d^{(\pm)}_m\delta_{n,m-1} -
d_m^{(\pm)} e_\alpha \delta_{n,m}]  \;.\label{lasg}\ea
From the brackets [ ] on the right hand side of (\ref{lasg})
we can construct infinitesimal $SU(2)$ gauge transformations analogous
to (\ref{pmsu}).  In this way
 $\ell^{(-)}_n $ generate the Lie-Poisson transformations.
We note  using (\ref{lA}),
 that $\ell^{(-)}_n$ contain the generators $G_n$   of the canonical
transformations (\ref{ct}) at first order in the expansion parameter
$1/\kappa$.  Thus the Lie-Poisson transformations (\ref{pm}) correspond
to a deformation of the canonical symmetries of the nonabelian Toda
lattice.

Just as $SU(2)$ transformations  are generated
the $SB(2,C)$ matrices,
$SB(2,C)$ transformations analogous to (\ref{pmsu}) are generated
the $SU(2)$ matrices $B_n$.

\section{Lattice Dynamics}

 It remains to write down the lattice
Hamiltonian associated with the Poisson brackets (\ref{lkm}),
 (\ref{dmdp}) and (\ref{dpdp}).
 It  should give $H_{1/\xi}$ in the continuum limit.  We shall
also require it to reduce to the nonabelian Toda lattice Hamiltonian
(\ref{tlH}) when $\xi\rightarrow \infty$.     Both of
these requirements are satisfied for
\Eq{ H^{lat}_{1/\xi} = -\frac1{4a\chi^2} \sum_n
\biggl((\xi^2+1)\; \Tr\;d^{(-)}_n {d^{(-)}_n}^\dagger -   \;
 2 \;\Tr\; (d^{(-)}_n +{d_n^{(-)}}^\dagger  )\biggr) \;.\label{lH}}
The Toda Hamiltonian (\ref{tlH}) is recovered using
\ba d^{(-)}_n &\rightarrow & B_n \;,\cr    & & \cr
\Tr \;d^{(-)}_n
{d^{(-)}_n}^\dagger &\rightarrow & -\frac 2{\kappa^2\chi^4}\Tr\; G_{n}^2
  +2 \;, \quad {\rm as}\; \kappa\rightarrow \infty\;,  \ea
while the continuum Hamiltonian (\ref{ham2}) is recovered using
\ba \Tr\;(d^{(-)}_n +{d^{(-)}_n}^\dagger ) &\rightarrow &
\frac {2 a^2}{\kappa^2} \biggl( \tilde I_\alpha(x_n)
\tilde I_\alpha(x_n)  -  \tilde J_\alpha(x_n)
\tilde J_\alpha(x_n)  \biggr) \;,\cr  & & \cr
\Tr \;d^{(-)}_n  { d^{(-)}_n}^\dagger &\rightarrow &
\frac {4 a^2}{\kappa^2}  \tilde I_\alpha(x_n)  \tilde I_\alpha(x_n)
 \;, \quad {\rm as}\; a\rightarrow 0\;. \ea

In section 3, we saw that the  Toda lattice Hamiltonian,
$ H^{lat}_{0} $ was not invariant under the most
general  canonical transformation (\ref{ct}).  Similarly,
$ H^{lat}_{1/\xi} $ is not invariant under the
general  transformations (\ref{pm}) [where we are assuming
(\ref{wmvm})].   Rather, they are preserved
only under the action of the global $SU(2)$ subgroup.
 Invariance of the $\Tr\;d^{(-)}_n$ terms implies that
 $v_1^{(\pm)}=v_2^{(\pm)}= ...=v_N^{(\pm)}=v^{(\pm)}
\;, N $ being the total number of links, while invariance of the
$\Tr \;d^{(-)}_n { d^{(-)}_n}^\dagger$  terms implies that
 $v^{(-)}=v^{(+)}=v$ is in $SU(2)$.
  On the other hand, the term    $\Tr \;d^{(-)}_n
{ d^{(-)}_n}^\dagger$  is invariant under local $SU(2)$
transformations generated by $\ell^{(-)}_m $.
Therefore $\ell^{(-)}_m $ has zero Poisson brackets with the
quadratic term in the Hamiltonian.  Its Hamilton equations of motion
are then determined by the linear terms
in (\ref{lH}) using the Poisson brackets (\ref{lasg}).  We find \Eq{
  a\chi^2\kappa\; \dot \ell^{(-)}_n    { \ell^{(-)}_n }^{-1} =
e^\alpha \Tr \;e_\alpha  (d^{(-)}_{n+1} -d^{(-)}_{n}
-{d^{(-)}_{n+1}}^\dagger +{d^{(-)}_{n}}^\dagger)   \;.}
From this property, we are unable  to construct the conserved charges
associated with the global symmetry purely from the $\ell^{(-)}_n$
matrices.  Thus we do not have the
analogue of the canonical generators $G$.

The equations of motion for $d^{(-)}_n$ are a bit more complicated.
From the Poisson brackets (\ref{lkm}) and
 (\ref{dmdp}) we find
\ba \pois{ d_n^{(-)} }{\sum_m \Tr\; d^{(-)}_m } &=& \frac 2\kappa
\biggl\{  d^{(-)}_n e^\alpha \;\Tr \;e_\alpha
  (d^{(-)}_{n+1} -d^{(-)}_{n}) + e_\alpha
d^{(-)}_n \;\Tr\; e^\alpha
  (d^{(-)}_{n} -d^{(-)}_{n-1})\biggr\}\;,\cr   & & \cr
 \pois{ d_n^{(-)} }{\sum_m \Tr\;
 { d^{(-)}_m}^\dagger } &=& \frac 2\kappa
\biggl\{ - d^{(-)}_n e^\alpha \;\Tr \;e_\alpha
  (d^{(-)}_{n+1} -d^{(-)}_{n})^\dagger + e^\alpha
d^{(-)}_n \;\Tr\; e_\alpha
  (d^{(-)}_{n} -d^{(-)}_{n-1})^\dagger \biggr\}\;,  \cr & & \cr
 \pois{ d_n^{(-)} }{\sum_m \Tr\;{ d^{(-)}_md^{(-)}_m}^\dagger } &=&
  \frac{2i} \kappa [d^{(-)}_n
{d^{(-)}_n }^\dagger - {d^{(-)}_{n-1}}^\dagger d^{(-)}_{n-1}
]_{t\ell}\;d^{(-)}_{n}\;.     \ea
  The equations of motion which follows from the Hamiltonian (\ref{lH})
  are then \ba  a\chi^2\kappa\; \dot d^{(-)}_n    { d^{(-)}_n }^{-1}
 &=& - \frac i2(1+\chi^4\kappa^2) \;[d^{(-)}_n
{d^{(-)}_n }^\dagger - {d^{(-)}_{n-1}}^\dagger d^{(-)}_{n-1}
]_{t\ell} \cr       & & \cr
& & +\;   e_\alpha   \;\Tr\; e^\alpha    (d^{(-)}_{n} -d^{(-)}_{n-1})
+   e^\alpha \;\Tr \; e_\alpha  (d^{(-)}_{n} -d^{(-)}_{n-1})^\dagger\cr
& &\cr    & &+\; d^{(-)}_n e^\alpha   {d^{(-)}_n}^{-1}
\;\Tr \;e_\alpha  (d^{(-)}_{n+1} -d^{(-)}_{n}
-{d^{(-)}_{n+1}}^\dagger +{d^{(-)}_{n}}^\dagger)   \;.\ea

\section{Application to two-dimensional Lattice Gauge Theories}

We saw above that the lattice descriptions of the chiral model
do not fully utilize the symmetries (be they canonical or Lie-Poisson)
of the Poisson brackets, as these are gauge symmetries.
The symmetry breaking was due to the presence of linear terms
in the two Hamiltonians  (\ref{tlH}) and  (\ref{lH}).
 On the other hand, if we only keep  quadratic-like
terms in the Hamiltonians, we can construct lattice gauge theories,
which is the purpose of this section.
 For the two different systems, the relevant gauge group
will be $SU(2)$, although it is implemented as canonical
transformations in one system
   and Lie-Poisson transformations in the other.

The restoration of the gauge symmetry (\ref{ct}) or (\ref{pm})
will mean that we will be left with a trivial theory because we can
 eliminate all but a few degrees of freedom.  With regard to the
canonically invariant theory defined by the Poisson brackets
(\ref{tlpbp}),  recall that $G_n$ appearing in
  (\ref{tlH}) are the generators of the gauge symmetries which here are
  implemented as canonical transformations
(\ref{ct}).  Therefore the Gauss law constraint
requires that we set \Eq{G_n=0\;,\label{gl}} at all lattice sites.
Furthermore, we can use the gauge symmetry (\ref{ct}) to eliminate
the degrees of freedom in $B_n$.   If we
 again assume  the periodic boundary conditions
 $B_{n+N}=B_n$ then this can be done everywhere except at one lattice
site.

Because of the constraint (\ref{gl})
 the quadratic term in the Hamiltonian
  (\ref{tlH}) vanishes.   To recover
 two dimensional lattice QCD  we need to
reexpress the nonabelian Toda theory in terms of a local algebra,
specifically   $[T^*SU(2)]^{\otimes N}$.  Here we associate a cotangent
   bundle of $SU(2)$ with each point on the lattice.
To see this we introduce a new set of traceless antihermitean matrices
 $E_n$ which play the role of the electric field along the links of
the lattice  and generate
right $SU(2)$ transformations at lattice site $n$.  Their  Poisson
brackets can be written       \ba  \frac 2{\chi^2}
\pois{\Enone}{\Emtwo} &=&  [C,\Emtwo ] \delta_{n,m} \;,\cr & & \cr
  \frac 2{\chi^2} \pois{\Enone}{\Bmtwo} &=&  -
\Bmtwo C\delta_{n,m} \;.   \label{pbfE} \ea  These equations
  along with the last equation in (\ref{tlpbp}) define the
 cotangent bundle of $SU(2)$ at each lattice site $n$.
  For a given $n$, $E_n$ and $B_n$ then span the six-dimensional
 phase space   of the rigid rotor.
To recover the first two   equations in (\ref{tlpbp})
we can set
 \Eq{G_n=E_{n}- B_{n+1} E_{n+1} B_{n+1}^\dagger\;.\label{GiEB}}
The electric fields
 $E_n$ undergo the following canonical transformations
\Eq{ E_n \rightarrow E'_n= v_n^{-1} E_n v_n \;,\label{ctfE}}
which together with   (\ref{ct}) preserve the Poisson brackets
 (\ref{pbfE}).    From the definition  (\ref{GiEB})  of $G_n$,
the Poisson brackets of $G_n$ with $E_m$ are given by
\Eq{  \frac 2{\chi^2}
\pois{\Enone}{\Gmtwo} =  [C,\Emtwo ] \delta_{n,m} \;,}
from which it again follows that
 $G_n$ are the generators of the canonical  transformations.
spanning   $[T^*SU(2)]^{\otimes N}$.

From (\ref{gl}) and (\ref{GiEB}),
the electric fields $E_n$ are subject to the constraints
 (\ref{gl}) [where $G_n$ is expressed in terms of $E_n$ using
(\ref{GiEB})].     The two
dimensional version of the Kogut-Susskind formulation\cite{ks}
of lattice QCD is recovered
 by choosing the Hamiltonian to be \Eq{H_{ks}=\sum_{n=1}^N\Tr\; E_n^2\;.}

With regard to the Lie-Poisson invariant theory defined by
Poisson brackets  (\ref{lkm}),  (\ref{dmdp}) and (\ref{dpdp}),
 if we restrict the gauge group to $SU(2)$, then
the analog of the Gauss law constraint is
 \Eq{  \ell^{(-)}_n=\BI\;,\label{dgl}
} at all lattice sites.    Using
(\ref{lA}) we recover (\ref{gl}) from (\ref{dgl})
in the limit $\kappa \rightarrow \infty$.
The remaining degrees of freedom in  $d^{(-)}_n $
can be eliminated using the gauge transformation
(\ref{pmsu}) [except at one lattice site, if we assume the periodic
boundary conditions  $d^{(-)}_{N+n} =  d^{(-)}_n $].

  Concerning the Hamiltonian for this system,
if we want to make a connection with two dimensional
QCD,    quadratic terms  like those appearing
 in (\ref{lH}) are unsuitable.  This is
  because they can be expressed solely in terms of the generators
$\ell^{(-)}_n$ of the Lie-Poisson transformation (\ref{pmsu}).  For this
we note that \Eq{\Tr \;d^{(-)}_n   { d^{(-)}_n}^\dagger
=\Tr \;\ell^{(-)}_n   {\ell^{(-)}_n}^\dagger \;.\label{ddH}}
Consequently such terms are trivial due to (\ref{dgl}).
As was true in the Kogut-Susskind theory, we can express the dynamical
 variables, here $\ell^{(-)}_n$ and $B_n$, in terms
of  variables which span a local algebra.
We can also write the Hamiltonian in terms of these variables.
The resulting system is a deformation of
the Kogut-Susskind formulation of gauge theories and has been examined
 previously in refs. \cite{fro},\cite{BSV}.

In analogy to (\ref{GiEB}), we set
\Eq{\ell^{(-)}_n =   k^{(-)}_n \tilde  k^{(-)}_{n+1}  \; ,}
where $  k^{(-)}_n$ and $ \tilde  k^{(-)}_{n}$ are $SB(2,C)$ matrices,
which are analogous to the electric fields $E_n$ in the Kogut-Susskind
 system.  The algebra (\ref{lbpb}) for
$\ell^{(-)}_n $ and $B_n$ is recovered from the local algebra
 \ba
  \{\knone,\kmtwo\} &=& [r,\knone \kmtwo ]\delta_{n,m}\;, \cr
  \{\ktnone,\ktmtwo\} &=& [r,\ktnone \ktmtwo ]\delta_{n,m}\;, \cr
  \{\knone,\ktmtwo\} &=& 0\;, \cr
  \{\knone,\Bmtwo\} &=&-\Bmtwo r\knone\delta_{n,m}\;,\cr
  \{\ktnone,\Bmtwo\} &=&\ktnone r\Bmtwo  \delta_{n,m}\;.  \label{tla}
\ea
If we also assume the Poisson brackets \Eq{
  \{\kpnone,\kmtwo\}  = [\tilde r,\kpnone \kmtwo ]\delta_{n,m} \;,}
where $k^{(+)}_n={k^{(-)\dagger}_n}^{-1}$, then we can show that
   $$\Tr \;k^{(-)}_n   { k^{(-)}_n}^\dagger $$
is gauge invariant, i.e. it has zero Poisson brackets with the
gauge generators    $\ell^{(-)}_n $.   Unlike (\ref{ddH}), it is
not trivial due to the constraints and it can be taken to be the
Hamiltonian of the system.   If we further set
 \Eq{k_n^{(-)}=\exp{\biggl\{
-\frac{2i}\kappa E_{n,\alpha}e^\alpha}\biggr\}\;,}
we can recover (up to factors and an infinite additive constant)
the Kogut-Susskind Hamiltonian $H_{ks}$
in the $\kappa\rightarrow\infty$ limit.  The resulting deformation
of the Kogut-Susskind formulation of lattice gauge theories was
examined previously (in two, three and four dimensions) in refs.
\cite{fro},\cite{BSV}.

Above we have seen that by writing
$d^{(-)}_n =B_n   k^{(-)}_n \tilde  k^{(-)}_{n+1}$, the non local
algebra of \cite{afs} can be expressed entirely in terms of a local
algebra.  The latter is given by
(\ref{tla}), along with the first Poisson bracket in (\ref{lbpb}).
From \cite{mss}, we have that
$(B_n , k^{(-)}_n)$ and $(B_n, \tilde  k^{(-)}_{n})$ define two
different paramtrizations of the classical double algebra (for each $n$),
one where the double variable is written as  the product
$B_n  k^{(-)}_n$, and the other where
 the double variable is written as  the product
 $ \tilde  k^{(-)}_{n}B_n$.

\section{Conclusion}

We have shown how to apply the  current algebra of ref. \cite{afs}
in order to get a new lattice description of the chiral
model.   As this current algebra admitted Lie-Poisson symmetries,
so did the new lattice description  of the chiral model.
  The  Lie-Poisson symmetries get promoted to quantum group symmetries
upon quantization.   The quantum mechanical commutation relations
for the operators analogous to $d^{(-)}_n$ are known\cite{afs}, while
those for
 the operators analogous to $d^{(+)}_n$ are readily obtained from
their Poisson brackets.  To get the quantum mechanical Hamiltonian
one basically only needs to replace the traces in
(\ref{lH}) with  deformed traces.  (See for example \cite{mss}.)

 The above lattice description of  the chiral model
  was possible because  the continuum limit of the lattice
  current algebra is the same as the algebra
 appearing in the alternative Hamiltonian description
of ref. \cite{Raj}, i.e. it is the $SL(2,C)$ Kac-Moody algebra.   In
refs. \cite {RSV},\cite{RSV2} this alternative
 Hamiltonian description  was generalized
to the case of the chiral model with a Wess-Zumino term
(whose coefficient was arbitrary).
The latter also relied upon the $SL(2,C)$ Kac-Moody algebra, only
here, unlike in (\ref{kma}), both central terms were required.
To include the Wess-Zumino term, we   generalize (\ref{kma})
 to\footnote{Now in comparing with
 ref. \cite {RSV}, the parameters $\rho$ and $\tau$ of
that reference are given by $\rho=-\xi\xi'$ and $\tau^2=-\xi^{-2}$.}
 \ba \{ \tilde I_\alpha(x), \tilde I_\beta(y)\}
  &=&-\{\tilde J_\alpha(x), \tilde J_\beta(y)\}
  =\epsilon_{\alpha \beta \gamma} \tilde I_\gamma(x) \delta(x-y)+
 \xi'\epsilon_{\alpha \beta \gamma} \tilde J_\gamma(x) \delta(x-y)\;,\cr
& & \cr   \{ \tilde I_\alpha(x), \tilde J_\beta(y)\} &=&
\epsilon_{\alpha \beta \gamma} \tilde J_\gamma(x) \delta(x-y)
-\xi'\epsilon_{\alpha \beta \gamma} \tilde I_\gamma(x) \delta(x-y)
+\frac{\xi}{2\chi^2}
 \delta_{\alpha\beta}\partial_x \delta(x-y) \;,\label{kmap}\ea
where $\xi'$ is real.  If we now define the complex current $j(x)$
according to   \Eq{j(x)={{\tilde I_\alpha(x)\sigma_\alpha +
 i\tilde J_\alpha(x)\sigma_\alpha}\over{1-i\xi'} }
 \;,} as opposed to (\ref{jox}), we recover the algebra given by
(\ref{jj}) and (\ref{jjdag}) where  $\kappa$ is now complex:
\Eq{\kappa = \frac \xi{\chi^2}(1-i\xi') \;.}  This
is the most general $SL(2,C)$ Kac-Moody algebra.  If we want
to obtain it in
  the continuum limit of the lattice current algebra
we need $\kappa$ in (\ref{lkm})  to be
complex.  Thus far, we have  not been successful in finding a consistent
algebra for this case, i.e. one that satisfies (\ref{inv}), and we
therefore have been unable to generalize our system
to get a new lattice description of the chiral
model with a Wess-Zumino term.

 Of course another concern is the question of integrability.
The conserved charges for the two dimensional
chiral model are well known.  Upon going to the lattice,
 a Lax pair  construction can be made
using the Toda model description if one works with the general linear
  group, rather than say $SU(2)$.\cite{fm}
However,   this construction is not readily adaptable  to other
cases.  It may be that neither of the models presented here are
integrable.  However, the Hamiltonian systems we examined on the
lattice are not unique, and further study may yield solvable models.

{\parindent 0cm{\bf Acknowledgements:}}
 A.S. was supported in part
by the Department of Energy, USA, under contract number
DE-FG05-84ER40141.

\end{document}